\def\simlt{\mathrel{\spose{\lower 3pt\hbox{$\mathchar''218$}}
     \raise 2.0pt\hbox{$\mathchar''13C$}}}
\def\simgt{\mathrel{\spose{\lower 3pt\hbox{$\mathchar''218$}}
     \raise 2.0pt\hbox{$\mathchar''13E$}}}
\documentstyle[aaspp4,epsf]{article} 
\begin{document}
\def\gtorder{\mathrel{\raise.3ex\hbox{$>$}\mkern-14mu
             \lower0.6ex\hbox{$\sim$}}}
\def\ltorder{\mathrel{\raise.3ex\hbox{$<$}\mkern-14mu
             \lower0.6ex\hbox{$\sim$}}}

\def\today{\number\year\space \ifcase\month\or  January\or February\or
        March\or April\or May\or June\or July\or August\or
        September\or
        October\or November\or December\fi\space \number\day}
\def\fraction#1/#2{\leavevmode\kern.1em
 \raise.5ex\hbox{\the\scriptfont0 #1}\kern-.1em
 /\kern-.15em\lower.25ex\hbox{\the\scriptfont0 #2}}
\def\spose#1{\hbox to 0pt{#1\hss}}
\def\rsun{R_{\odot}}
\def\msun{M_{\odot}}
\def\rhosun{\rho_{\odot}}
\def\rstar{R_{*}}
\def\mstar{M_{*}}
\def\rplanet{R_{p}}
\def\mplanet{M_{p}}
\def\mp{M_p}
\def\rp{R_p}
\def\sqdepth{\sqrt{\Delta F}}
\def\tf{t_F}
\def\tt{t_T}
\def\rhostar{\rho_*}

\title{On the Unique Solution of Planet and Star Parameters from an Extrasolar Planet Transit Light Curve}

\author{S.\ Seager\footnote{Institute for Advanced Study, Einstein
Drive, Princeton, NJ, 08540; seager@ias.edu} and
G.\ Mall\'en-Ornelas\footnote{Princeton University Observatory,
Peyton Hall, Princeton NJ, 08544-0001 and Departamento de
Astronom\'{\i}a y Astrof\'{\i}sica, Pontificia Universidad Cat\'olica
de Chile, Casilla 306, Santiago 22, Chile; mallen@astro.princeton.edu}}

\pagestyle{plain}

\begin{abstract}

There is a unique solution of the planet and star parameters from a
planet transit light curve with two or more transits if the planet has
a circular orbit and the light curve is observed in a band pass where
limb darkening is negligible. The existence of this unique solution is
very useful for current planet transit surveys for several reasons.
First, there is an analytic solution that allows a quick parameter
estimate, in particular of $R_p$. Second, the stellar density can be
uniquely derived from the transit light curve alone. The stellar
density can be used to immediately rule out a giant star (and hence a
much larger than planetary companion) and can also be used to put an
upper limit on the stellar and planet radius even considering slightly
evolved stars. Third, the presence of an additional fully blended star
that contaminates an eclipsing system to mimic a planet transit can be
largely ruled out from the transit light curve given a spectral type
for the central star. Fourth, the period can be estimated from a
single-transit light curve and a measured spectral type. All of these
applications can be used to select the best planet transit candidates
for mass determination by radial velocity follow-up. To use these
applications in practice, the photometric precision and time sampling
of the light curve must be high (better than 0.005~mag precision and
5~minute time sampling).
\end{abstract}

\section{Introduction}

Planet transit searches promise to be the next big step forward for
extrasolar planet detection and characterization.  Every transiting
planet discovered will have a measured radius, which will provide
constraints on planet composition, evolution, and migration
history. Together with radial velocity measurements, the absolute mass
of every transiting planet will be determined.  Transiting planets can
be discovered around distant stars and in a variety of
environments. Due to their special geometry many follow-up
observations of transiting planets are possible, such as atmosphere
transmission spectroscopy (note the first extrasolar planet atmosphere
detection by Charbonneau et al.\ 2002), search for moons and rings
(Brown et al.\ 2001), and detection of oblateness and the
corresponding constraint on rotation rate (Seager \& Hui
2002). Although no planet candidates discovered by the transit method
have yet been confirmed by mass measurements, many searches are
currently ongoing. The OGLE-III planet search (Udalski et al.\ 2002)
has observed numerous high-precision transit light curves from objects
with small radii, including several potential planets.  The EXPLORE
search (Mall\'en-Ornelas et al.\ 2002; Yee et al.\ in preparation) has
four potential planet candidates based on both photometric light
curves and follow-up radial velocity measurements (Mall\'en-Ornelas et
al., in preparation).  The Vulcan planet search (Borucki et al. 2001)
has some published results on transit candidates that, with radial
velocity measurements, were determined to be eclipsing binary stars
(Jenkins, Caldwell, \& Borucki 2002).

Follow-up mass determination by radial velocity measurements are
needed for planet transit candidates because late M dwarfs ($M
\geq 80 M_J$), brown dwarfs ($13 M_J < M < 80 M_J$), and gas giant
planets ($M \leq 13 M_J$) are all of similar sizes. This is due to
a coincidental balance between Coulomb forces (which cause $R \sim
M^{1/3}$) and electron degeneracy pressure (which causes $R \sim
M^{-1/3}$). A high yield of confirmed planets from a list of
planet candidates considered for mass follow-up is important,
especially for planet searches with faint stars (e.g., fainter
than 13th magnitude) which require relatively long exposures on
8-m class telescopes (e.g., 20--30 mins per star for a single
radial velocity measurement). Hence understanding the transit
light curves before follow-up can be crucial if a given project
has a large number of planet transit candidates.

An analytical solution is always worthwhile to understand the general
properties of a given physical system in an intuitive form. This is
the case even when numerical fits are in practice the best way to
determine the system parameters. In the case of a planet transit light
curve we found that there is a unique solution for the five parameters
stellar mass $M_*$, stellar radius $R_*$, companion radius $R_p$,
orbital distance $a$, and orbital inclination $i$, under some
important assumptions. This unique solution has several interesting
applications --- especially when the photometric precision and time
sampling are high --- including selection of the best planet
candidates for follow-up mass measurements. Selection of the best
candidates is especially important if the tendency for planets with
small orbital distances to have low mass (Zucker \& Mazeh 2002) ---
and hence low radial velocity amplitudes --- is generally true.  In
this case, on average, more effort to detect the planet mass via
radial velocity variations of the parent star will be needed since the
primary star's radial velocity amplitude scales linearly with planet
mass.

The unique solution to a light curve with two or more transits was
first mentioned in Mall\'en-Ornelas et al.\ (2002) where an
approximate set of equations and a short description were
presented. Sackett (1995) briefly touches on the unique solution by
outlining parameter derivation with a known stellar spectral type,
including a mention of period determination from a single transit. We
begin this paper by describing the assumptions necessary to determine
the unique solution in \S2. In \S3, for the first time, we present
both the general set of equations that describe a planet transit light
curve and their analytic solution. We discuss the errors in the
parameters, and hence the limiting photometric precision and time
sampling needed for the applications outlined in this paper in
\S4. The complications of limb-darkening are discussed in \S5. In \S6
we present four interesting applications that are made possible by the
unique solution to the planet transit light curve. \S7 concludes this
paper with a summary.

\section{Assumptions}
\label{sec-assumptions} The unique determination of the stellar
mass $M_*$, stellar radius $R_*$, companion radius $R_p$, orbital
distance $a$, and orbital
inclination $i$ from a light curve with two or more eclipses requires the following assumptions:\\
$\bullet$ The planet orbit is circular; \\
$\bullet$ $M_p \ll M_*$ and the companion is dark compared to the central star;\\
$\bullet$ The stellar mass-radius relation is known;\\
$\bullet$ The light comes from a single
star, rather than from two or more blended stars.\\
The unique determination also requires the light curve to fulfill the following conditions:\\
$\bullet$ The eclipses
have flat bottoms  which implies that the companion is fully
superimposed on
the central star's disk;\\
$\bullet$ The period can be derived from the light curve (e.g.,
the two observed eclipses are consecutive).

The first three assumptions are all reasonable for current extrasolar
planet transit searches. Circular orbits are expected for short-period
planets due to their short tidal circularization timescale, and all
but one of the currently known short-period planets ($<4.3$ days) have
eccentricities consistent with zero\footnote{Extrasolar Planets
Encyclopaedia http://cfa-www.harvard.edu/planets/}.  All current
ground-based transit searches are searching for or expecting to find
mostly short-period planets which have the highest geometric
probability to show transits. The stellar mass-radius relation is
reasonably well-known for each separate class of stars (e.g., main
sequence), and as we show later, the mass-radius relation is not
needed for a derivation of all parameters. The only assumption that
has a significant chance of being wrong is the presence of a blended
star (e.g., from a physical stellar companion); this situation and its
possible identification are discussed in \S\ref{sec-applications}. The
required conditions listed above are also all reasonable.
Flat-bottomed transits will appear in a band pass where limb darkening
is negligible, such as $I$-band. Several transit surveys are using
$I$-band including OGLE-III (Udalski et al.\ 2002), EXPLORE
(Mall\'en-Ornelas et al.\ 2002), and STEPPS (C. J. Burke et al., in
preparation), and follow-up measurements for other surveys could be
taken in $I$-band or in even a redder color in order to exploit the
unique solution discussed in this paper. (See \S\ref{sec-ld} for a
discussion of limb darkening). For the rest of the paper we will work
under the assumptions and conditions listed above, unless otherwise
stated.

\section{The Equations and Solution for a Light Curve with Two or More Transits}

\subsection{The General System of Equations}
\begin{figure}
\plotone{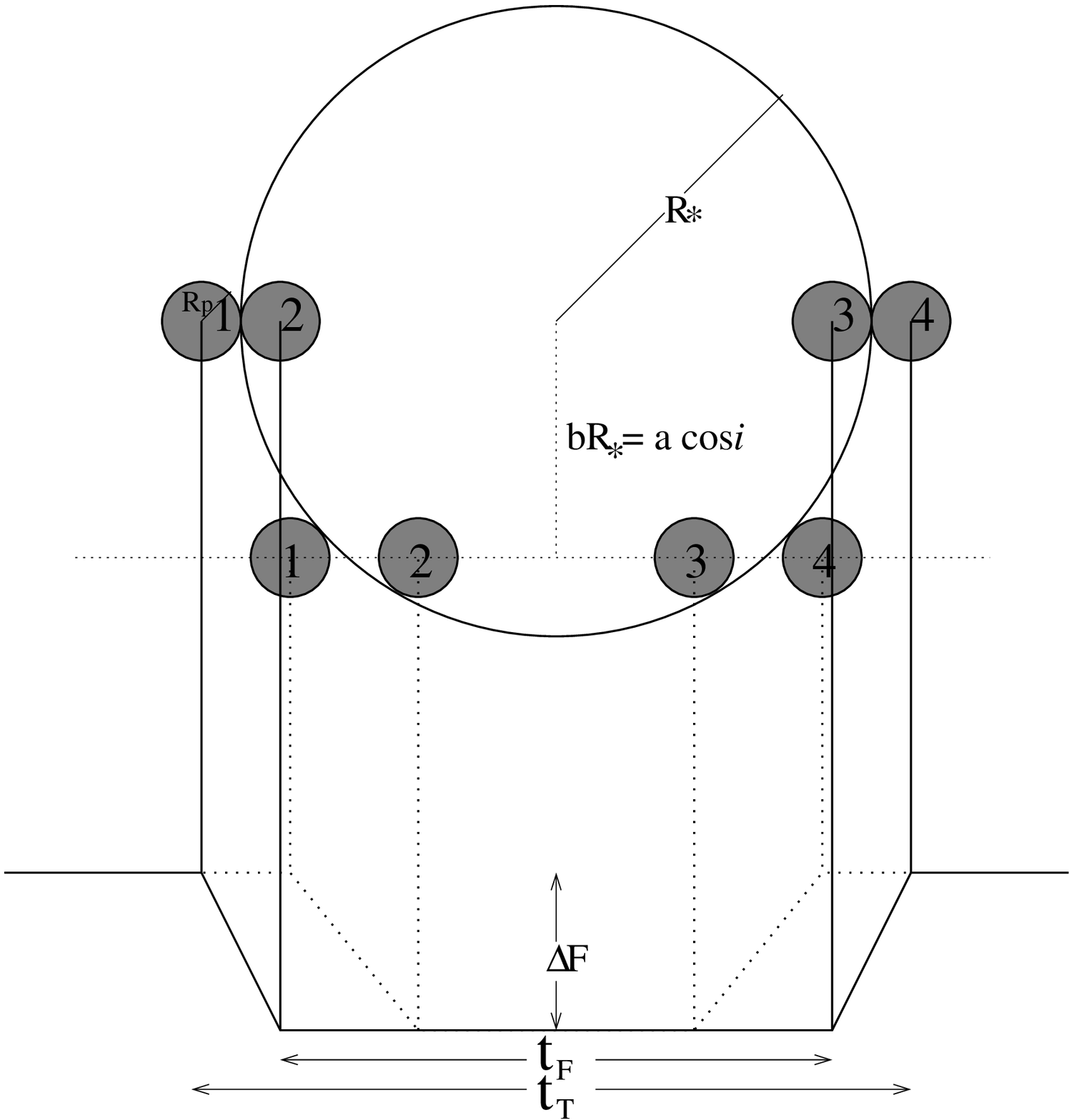}
\caption{Definition of transit light curve observables.  Two schematic
light curves are shown on the bottom (solid and dotted lines), and the
corresponding geometry of the star and planet is shown on the top.
Indicated on the solid light curve are the transit depth, $\Delta F$;
the total transit duration, $t_T$; and the transit duration between
ingress and egress, $t_F$, (i.e., the ``flat part'' of the transit light
curve when the planet is fully superimposed on the parent
star). First, second, third, and fourth contacts are noted for a
planet moving from left to right.  Also defined are $\rstar$, $\rp$,
and impact parameter $b$ corresponding to orbital inclination $i$.
Different impact parameters $b$ (or different $i$) will result in
different transit shapes, as shown by the transits corresponding to
the solid and dotted lines. }
\label{fig:schema} 
\end{figure}

There are five equations that completely describe the planet
transit light curve. The first three equations (equations
(\ref{eq:depthorig})--(\ref{eq:durationorig})) describe the
geometry of the transit in terms of transit depth, transit shape,
and transit duration (see Figure~\ref{fig:schema}). For a planet
transit light curve which is due to two spheres passing in front
of each other, the geometry is relatively straightforward (see
Sackett (1995) for a derivation of the the transit duration
equation~(\ref{eq:durationorig})). Here we parameterize the
transit shape by both $t_T$, the total transit duration (first to
fourth contact), and by $t_F$, the duration of the transit
completely inside ingress and egress (second to third contact).
The three geometrical equations that describe the transit light
curve depend on four observables: the period $P$, the transit
depth $\Delta F$, $t_F$, and $t_T$. See Figure~\ref{fig:schema}
for an illustrative definition of $\Delta F$, $t_F$, and $t_T$.
In addition to the three geometrical equations there are two
physical equations
(equations~(\ref{eq:Keplerorig})--(\ref{eq:MRorig})), Kepler's
Third Law and the stellar mass-radius relation. It is these
physical equations that break the degeneracy of the mathematical
description of two spheres passing in front of each other, by
setting a physical scale. It is this physical scale, together with
the geometrical description, that allows the unique solution.

The equations are: the transit depth, $\Delta F$, with $F$ defined as the
total observed flux,
\begin{equation}
\label{eq:depthorig} \Delta F \equiv \frac{F_{no \hspace{0.04in}
transit} - F_{transit}}{F_{no \hspace{0.04in} transit}} =
\left(\frac{\rplanet}{\rstar}\right)^2;
\end{equation}
the transit shape, described by the ratio of the duration of the
``flat part''of the transit ($t_F$) to the total transit duration ($t_T$)
\begin{equation}
\label{eq:shapeorig}
\frac{ t_F}{t_T} = \frac{
 \arcsin {\left( \frac{\rstar}{a}\left[
\frac{\left(1 - \frac{\rp}{\rstar}\right)^2 - \left(\frac{a}{\rstar} \cos i\right)^2}{1 - \cos^2 i}\right]^{1/2}\right)}}
{
 \arcsin {\left( \frac{\rstar}{a}\left[
\frac{\left(1 + \frac{\rplanet}{\rstar} \right)^2 - \left(\frac{a}{\rstar} \cos i\right)^2}{1 - \cos^2 i}\right]^{1/2}\right)}};
\end{equation}
the total transit duration
\begin{equation}
\label{eq:durationorig} t_T = \frac{P}{\pi}\arcsin {\left(
\frac{\rstar}{a}\left[ \frac{\left(1 + \frac{\rplanet}{\rstar}
\right)^2 - \left(\frac{a}{\rstar} \cos i\right)^2}{1 - \cos^2 i}\right]^{1/2}\right)};
\end{equation}
Kepler's Third Law, assuming a circular orbit, where $G$ is the
universal gravitational constant and $M_p$ the planet mass,
\begin{equation}
\label{eq:Keplerorig}
P^2 = \frac{4 \pi^2 a^3}{G(\mstar + \mplanet)};
\end{equation}
and the stellar mass radius relation,
\begin{equation}
\label{eq:MRorig}
\rstar = k \mstar^x,
\end{equation}
where $k$ is a constant coefficient for each stellar sequence
(main sequence, giants, etc.) and $x$ describes the power law of
the sequence (e.g., $x \simeq 0.8$ for F--K main sequence stars
(Cox 2000)).

\subsection{Analytical Solution}
\label{sec-soln}
\subsubsection{Four Parameters Deriveable from Observables}
\label{sec-foursoln}
We ultimately wish to solve for the five unknown parameters
$\mstar$, $\rstar$, $a$, $i$, and $\rp$ from the five equations above. 
It is first useful to note that four combinations of
physical parameters can be found directly from the
observables ($\Delta F$, $t_T$, $t_F$, and $P$) using only the
first four equations above (the three transit geometry equations
and Kepler's Third Law with $M_p \ll \mstar$); this avoids any
uncertainty from the stellar mass-radius relation.

The four combinations of parameters are: the planet-star radius
ratio which trivially follows from equation~(\ref{eq:depthorig}),
\begin{equation}
\frac{\rp}{\rstar} = \sqdepth;
\label{eq:depthdef}
\end{equation}
the impact parameter $b$, defined as the projected distance between the
planet and star centers during mid-transit in units of $\rstar$
(see Figure~\ref{fig:schema}), and which can be derived directly
from the transit shape equation~(\ref{eq:shapeorig}), together with 
equation~(\ref{eq:depthdef}),
\begin{equation} b
\label{eq:borig} \equiv \frac{a}{\rstar}\cos i  = \left[ \frac{ (1
- \sqrt{\Delta F})^2
 - \frac{ \sin^2 \frac{t_{F} \pi}{P}}
{\sin^2 \frac{t_{T} \pi}{ P}} (1 + \sqrt{\Delta F})^2}
{1 - \frac{ \sin^2 \frac{t_{F} \pi}{P}}
{\sin^2 \frac{t_{T} \pi}{ P}}}
\right]^{1/2};
\end{equation}
the ratio $a/ \rstar$ which can be derived directly from the
transit duration equation~(\ref{eq:durationorig}),
\begin{equation}
\label{eq:arorig} \frac{a}{R_*} = \left[ \frac{ (1 + \sqrt{\Delta
F})^2 - b^2 (1 - \sin^2 \frac{t_{T} \pi}{P})} { \sin^2
\frac{t_{T} \pi}{P}} \right]^{1/2};
\end{equation}
and the stellar density $\rhostar$ which can be derived from the
above equation for $a/\rstar$ and Kepler's Third Law with $M_p \ll
\mstar$ (equation~(\ref{eq:Keplerorig})),
\begin{equation}
\label{eq:rhoorig} \frac{\rhostar}{\rho_{\odot}} \equiv
\frac{\mstar/ \msun}{(\rstar/\rsun)^3}= \left[ \frac{4 \pi^2}{P^2
G} \right] \left[\frac{ (1 + \sqrt{\Delta F})^2 - b^2 (1 - \sin^2
\frac {t_T \pi}{P}) }{ \sin^2 \frac {t_T \pi}{P}}\right]^{3/2}.
\end{equation}
The parameters $b$ and $a/\rstar$ are dimensionless. The density
can also be written with the first term on the right hand side of
equation~(\ref{eq:rhoorig}) replaced by $\frac{4 \pi^2}{P^2 G} =
\frac{365.25^2}{P^2 215^{3}}$, with $P$ in days.

It is interesting to consider the geometrical and physical origin
of these combinations of parameters. The impact parameter $b$
depends almost entirely on the transit shape (parameterized by
$t_F/t_T$) and ratio of planet and star sizes ($\sqdepth$). To a
lesser extent (cf. \S\ref{sec-simpsoln}) $b$ depends mildly on the
period.  The term $a/\rstar$ is the ratio of orbital distance to
planet radius; to first order it is related to the ratio of
transit duration to total period. The term $a/\rstar$ is also
dependent on the impact parameter $b$ and planet-star size ratio,
because these parameters affect the transit duration. The stellar
density, $\rhostar$, comes from Kepler's Third Law and the
transit duration $t_T$; Kepler's Third Law describes how much
mass is enclosed inside the planet's orbit and the stellar radius
is described by the transit duration with a physical scale set by
Kepler's Third Law. Again, $\rhostar$ is also dependent on the
impact parameter $b$ and the planet-star size ratio, because
these parameters affect the transit duration.

\subsubsection{The Five Physical Parameters}
\label{sec-fivesoln}
The five physical parameters $R_*$, $M_*$, $i$, $a$, and $R_p$ can
be derived from the above solution for $\rp/ \rstar$, $b$,
$a/R_*$, and $\rhostar$ by using one additional equation: the
stellar mass-radius relation (equation~(\ref{eq:MRorig})). To
derive $\mstar$, consider equation~(\ref{eq:rhoorig}) together
with the stellar mass-radius relation in the form $\rhostar/
\rhosun \equiv \mstar/ \msun \left(\rstar / \rsun \right)^{-3}
 = \left(\mstar / \msun \right)^{1-3x} 1/k^3$:
\begin{equation}
 \label{eq:morig}
 \frac{\mstar}{\msun} = \left[ k^3 \frac{\rhostar}{\rhosun} \right]^{\frac{1}{1-3x}}.
\end{equation}
The stellar radius can be derived from the stellar mass by the
stellar mass-radius relation, or from the density directly,
\begin{equation}
\label{eq:rorig}
\frac{\rstar}{\rsun} = k \left(\frac{\mstar}{\msun}\right) ^x =
\left[k^{1/x} \frac{\rhostar}{\rhosun}\right]^{\frac{x}{(1-3x)}};
\end{equation}
the orbital radius $a$ can be derived from $M_*$ and from
Kepler's third law with $M_p \ll M_*$,
\begin{equation}
\label{eq:aorig}
 a = \left[ \frac{P^2 G \mstar} {4
\pi^2}\right]^{1/3};
\end{equation}
based on the definition of impact parameter (equation~(\ref{eq:borig})), the orbital inclination is,
\begin{equation}
\label{eq:iorig}
 i = \cos^{-1}\left(b \frac{\rstar}{ a}\right),
\end{equation}
and most importantly the planetary radius is
\begin{equation}
\label{eq:rporig}
 \frac {\rp}{\rsun} = \frac{\rstar}{\rsun} \sqdepth = \left[k^{1/x} \frac{\rhostar}{\rhosun}\right]^{\frac{x}{(1-3x)}} \sqdepth.
\end{equation}
For main sequence stars $k=1$ and $x \approx 0.8$, in which case
$\frac{\rp}{\rsun} = \left(\frac{\rhostar}{\rhosun}\right)^{-0.57} \sqdepth $.

\subsection{The Simplified Set of Equations and Their Solution}
\label{sec-approx} 
\begin{figure}
\plotone{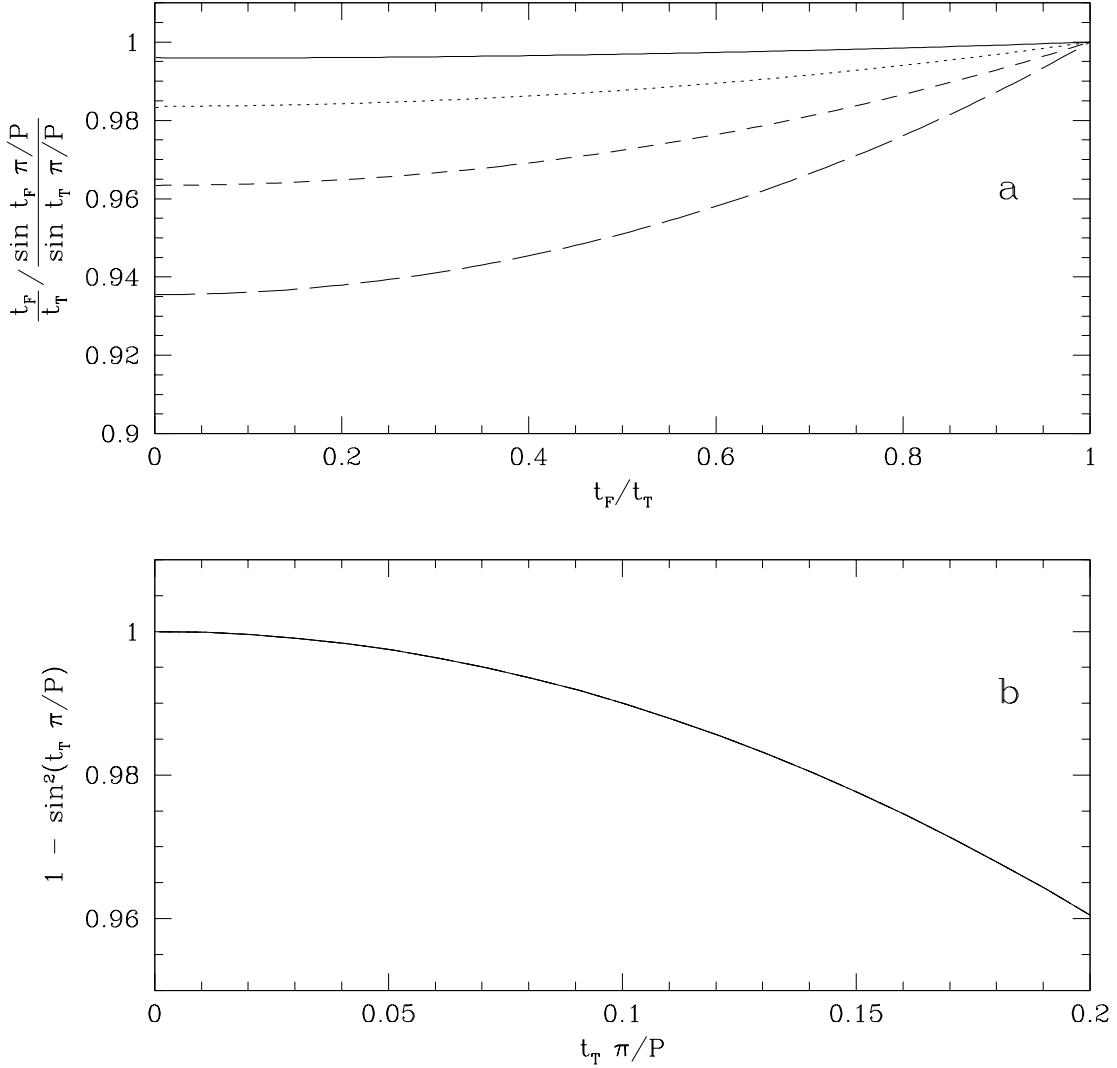} 
\caption{The validity of the approximation $t_T P/\pi \ll 1$ (or
equivalently $\rstar \ll a$). This approximation allows $\sin t_T
P/\pi \approx t_T P/\pi$. Panel a: the term of interest $\frac{\sin
(t_F \pi/P)}{\sin( t_T \pi/P)}$ is reasonably well approximated by
$t_F/t_T$. The different lines correspond to different values of $t_T
\pi/P$: 0.05 (solid line), 0.1 (dotted line), 0.15 (short-dashed
line), 0.2 (long dashed line).  Panel b: the term of interest $ 1-
\sin^2(t_T \pi / P)$ compared to its value, 1, under the approximation
$t_T P/\pi \ll 1$. The short-period planet transit systems of interest
will have $t_T \pi/P < 0.15$ and usually $t_T \pi/P \lesssim 0.1$.}
\label{fig:sinapprox} 
\end{figure}

The equations and five-parameter solution take
on a simpler form under the assumption $\rstar \ll a$. This
assumption is equivalent to $t_T \pi/P \ll 1$ (from
equation~(\ref{eq:arorig})), and has as its consequence $\cos
i\ll 1$ (from equation~(\ref{eq:iorig})).  Systems we are
interested in generally have $t_T \pi /P < 0.15$ and likely
$t_T \pi /P \lesssim 0.1$ (or $\rstar /a \gtrsim 1/8)$.
Mathematically this assumption allows $\arcsin x \approx x$ and
$\sin x \approx x$. Under this approximation, $\frac{\sin t_F
\pi/P}{\sin t_T \pi/P} \approx t_F/t_T$. A comparison of these
two terms is shown Figure~\ref{fig:sinapprox}a; for cases of
interest the terms agree to better than 4\% and much better in most
cases. Under the approximation $t_T \pi/P \ll 1$, a second term
of interest, $1-\sin^2(t_T \pi/P) \approx 1$. A comparison of
this term as a function of $t_T \pi /P$
(Figure~\ref{fig:sinapprox}b) shows agreement to better than
2.5\% for cases of interest. The simplified
solution will allow us to explore useful applications
analytically in \S\ref{sec-applications}.

\subsubsection{The Simplified Equations}
 Under the approximations $t_T \pi/P \ll 1$,
the transit shape (equation~(\ref{eq:shapeorig})) with $\arcsin x
\approx x$ becomes independent of $P$,
\begin{equation}
\label{eq:shapeapprox} \left(\frac{t_F}{t_T}\right)^2 = \frac{ \left( 1
- \frac{R_p}{R_*}\right)^2 - \left(\frac{a}{R_*} \cos i\right)^2}
{\left( 1 + \frac{R_p}{R_*}\right)^2 - \left(\frac{a}{R_*} \cos
i\right)^2},
\end{equation}
and the total transit duration,
(equation~(\ref{eq:durationorig})), with $\cos i \ll 1$ becomes
\begin{equation}
\label{eq:length} \label{eq:durationapprox} t_T = \frac{P \rstar
}{\pi a}\sqrt{\left(1 + \frac{\rp}{\rstar}\right)^2 -
\left(\frac{a}{R_*} \cos i\right)^2}.
\end{equation}
 The other three equations for transit depth
(equation~(\ref{eq:depthorig})), Kepler's Third Law
(equation~(\ref{eq:Keplerorig})), and the mass-radius relation
(equation~(\ref{eq:MRorig})), remain the same as in the
exact solution.  Note that by substituting $b = \frac {a}{\rstar} \cos i$ and
$\sqdepth = \frac {\rp}{\rstar} $ the above equations take a very simple form.

\subsubsection{The Simplified Analytical Solution}
\label{sec-simpsoln} The solution to the simplified equations is
more useful than the exact solution for considering the general
properties of light curves because $P$ either cancels out of the
solution or is a simple factor which can cancel out in parameter
ratios. The impact parameter $b$ (equation~(\ref{eq:borig})),
under the approximation $t_T \pi/P \ll 1$, becomes
\begin{equation}
 \label{eq:bapprox}
  b = \left[ \frac{ (1 - \sqdepth)^2 -
 \left(\frac{t_F}{t_{T}}\right)^2 (1 + \sqdepth)^2}{1 -
 \left(\frac{t_F}{t_T}\right)^2} \right]^{1/2};
\end{equation}
the ratio $a/ \rstar$ (equation~(\ref{eq:arorig})) becomes
\begin{equation}
 \label{eq:arapprox}
 \frac{a}{R_*} = \frac{2P}{\pi} \frac{\Delta F ^{1/4}}
 {\left(t_T^2 - t_F^2\right)^{1/2}};
\end{equation}
and the stellar density $\rhostar$ (equation~(\ref{eq:rhoorig}))
becomes
\begin{equation}
\label{eq:rhoapprox} \frac{\rhostar}{\rhosun} = \frac{32}{G \pi} P
\frac{ \Delta F^{3/4}}{\left({t_T}^2 - t_F^2\right)^{3/2}}.
\end{equation}
Note that for $P, t_F$ and $t_T$ in days, the first factor on the
right side of equation~(\ref{eq:rhoapprox}) becomes $32 / G \pi = 3.46
\times 10^{-3}$.  The equation for $\rp / \rstar$
(equation~(\ref{eq:depthdef})) clearly remains the same as in the
non-simplified case. The stellar mass, stellar radius, orbital
distance, orbital inclination, and planet radius can be derived as
before with equations~(\ref{eq:morig})--(\ref{eq:rporig}).

\section{Errors}
\label{sec-errors}
In principle the unique solution of $\rhostar$ and of the parameters
$\mstar$, $\rstar$, $i$, $a$, and $\rp$ provides a powerful tool for
understanding the transit light curve and more importantly for
selecting the best transit candidates for radial velocity
follow-up. In practice the usefulness of the unique solution is
limited by errors caused by the limited photometric precision and time
sampling of real data. The errors in the the star-planet parameters
are very non-Gaussian and often correlated so a simulation is
necessary in order to estimate errors. To compute the errors as a
function of photometric precision and time sampling, we generated one
thousand simulated transits with added Gaussian noise in the
photometry for each of several combinations of photometric precision
($\sigma$) and time sampling ($\delta t$).  We considered values of
$\sigma$ of 0.0025, 0.005, 0.01, and 0.015 mag and values of $\delta
t$ of 2.7, 6, and 12 minutes. Note that the shortest time sampling and
highest photometric precision are reachable by current transit surveys
(e.g., Mall\'en-Ornelas et al. 2002). We then fit the simulated
transits for period, phase, depth, impact parameter, and stellar
density with a $\chi^2$ minimization fit. Additionally we solved for
the planet radius using the main sequence stellar mass-radius relation
with $x=0.8$. For specificity we chose a star-planet model of $P=3.0$
days, $\mstar =
\msun$, $\rstar=\rsun$, $\Delta F = 2$\% (hence $\rp=0.14 \rsun = 1.45
R_J$). Limb darkening was not included (see \S\ref{sec-ld}). We
considered models with two different impact parameters ($b=0.2$ which
corresponds to $t_F/t_T=0.74$ and $b=0.7$ which corresponds to
$t_F/t_T=0.55$), because the errors are very sensitive to impact
parameter of the input transit. Although for specificity we have
focused on a single model (with two different impact parameters), for
the purpose of error estimates changing $t_T$ (by a change in $P$,
$\Delta F$, $a$; see equation~(\ref{eq:durationorig}) or
(\ref{eq:durationapprox}) ) is equivalent to a linear change in time
sampling, and changing $\Delta F$ (by a change in $\rp$ or $\rstar$;
see equation~(\ref{eq:depthorig})) is equivalent to a linear change in
photometric precision. Thus errors in parameters from other models can
be considered using the same computational results. Here we focus on
the two most interesting parameters, $\rhostar$ which can tell us if
the star is on or close to the main sequence, and $\rp$ an obvious
quantity of interest for selecting the best planet candidates.

\begin{figure}
\plotone{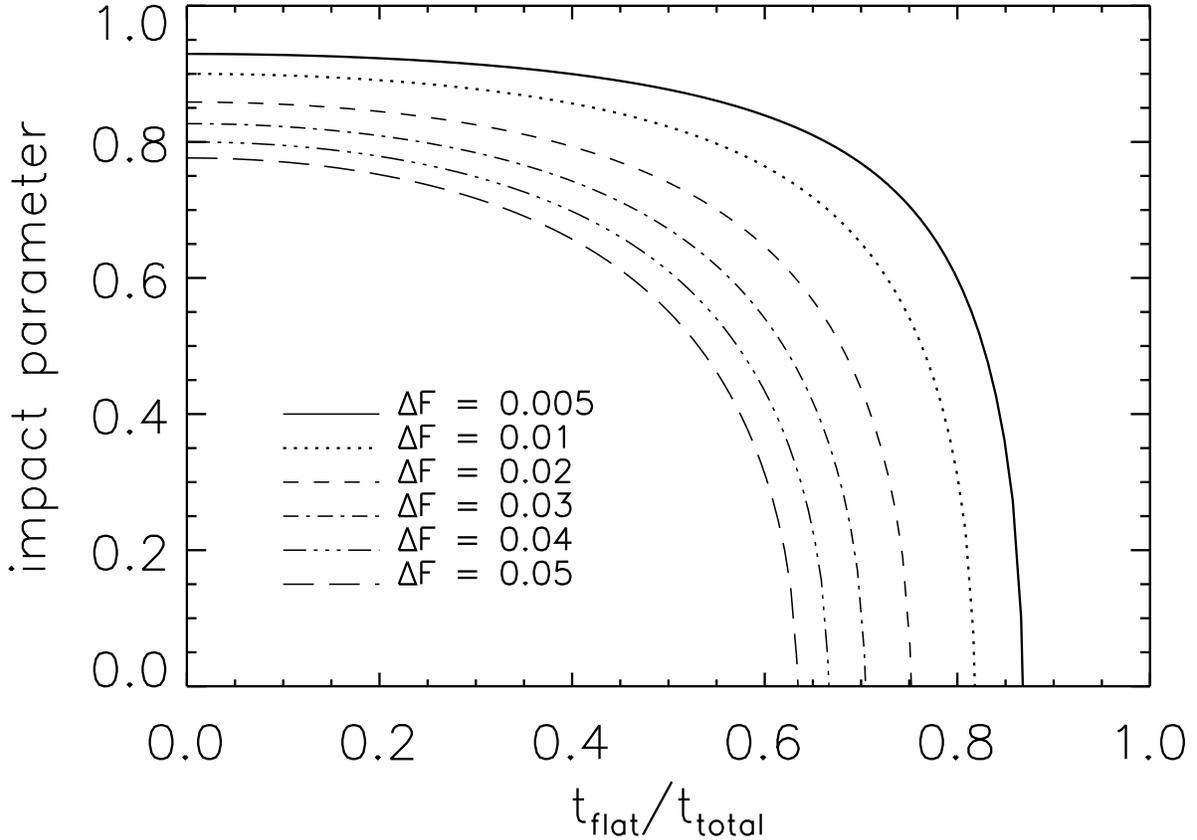} 
\caption{The one-to-one correspondence between the transit shape as
parameterized by $t_F/t_T$ and the impact parameter $b$ for a given
transit depth $\Delta F$ (equation~(\ref{eq:bapprox})). This
one-to-one correspondence is one of the elements that makes the unique
solution to the transit light curve possible. Transits with a given
$\Delta F$ can only fall along a given $b$--$t_F/t_T$ curve, and
for a given $\Delta F$ there is a maximum $b$ and a corresponding
minimum $t_F/t_T$. For large values of $t_F/t_T$ (box shaped
transits), a small change in $t_F/t_T$ can result in a large change in
$b$---making it difficult to derive $b$ accurately from the transit
light curve.  This effect also causes an underestimate in $b$ when the
transit light curve is noisy, because a symmetric error in $t_F/t_T$
causes a very asymmetric error in $b$.}
\label{fig:bvstf}
\end{figure}

The errors in $\rhostar$ and $\rp$ are dominated by errors in the
impact parameter $b$, which we discuss first.
Figure~\ref{fig:bvstf} shows that for a given $\Delta F$ there is a one-to-one
correspondence between the impact parameter $b$ and the transit
shape as parameterized by $t_F/t_T$; this one-to-one
correspondence is one of the reasons for the existence of the
unique solution. For ``box-shaped'' transits with large values of
$t_F/t_T$, a small change in $t_F/t_T$ can result in a very large
change in $b$---making it difficult to derive $b$ accurately from
box-shaped transit light curves. This is not the case for
transits with small $t_F/t_T$ (i.e., with very long
ingress/egress times), where $b$ changes little even for a
relatively large change in $t_F/t_T$. In the case of noisy data,
$b$ will be underestimated due to asymmetric errors resulting
from the non-linear relation between $b$ and $t_F/t_T$. An
underestimate in $b$ corresponds to an overestimate in $\rhostar$
and an underestimate in $\rstar$ and $\rp$.

\begin{figure}
\plotfiddle{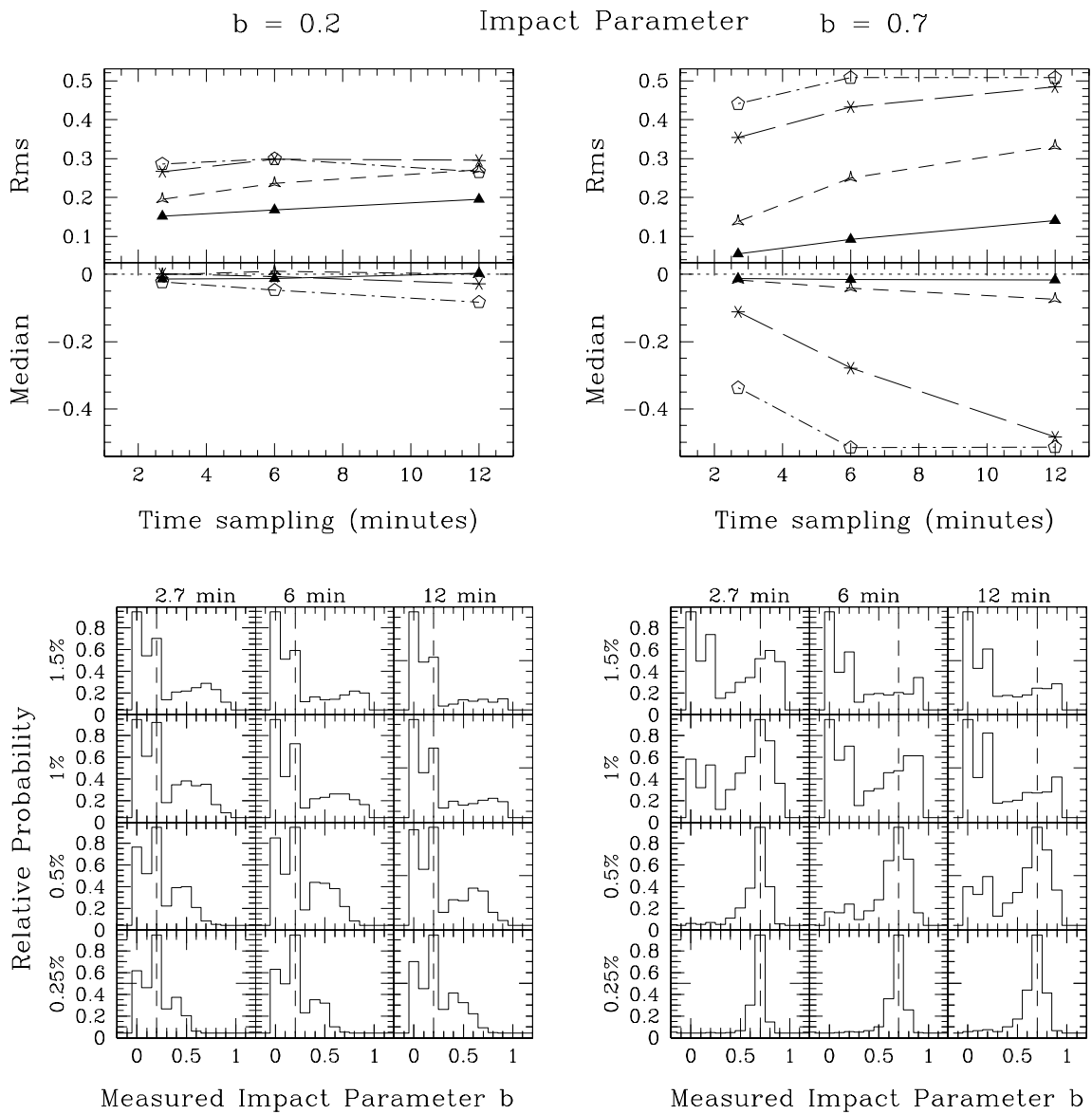}{5in}{0}{80}{80}{-250}{-120}
\caption{Errors in the derived impact parameter for the model $P=3.0$
days, $\mstar = \msun$, $\rstar=\rsun$, $\Delta F = 2$\% (hence
$\rp=0.14 \rsun = 1.45 R_J$), and no limb darkening.  The left panels
are for $b=0.2$ and the right panels for $b=0.7$.  For each
combination of photometric precision $\sigma = 0.0025$, 0.005, 0.01
and 0.015 mag and time sampling $\delta t = 2.7$, 6 and 12 min, 1000
noisy model transits were created and fit.  The top panels show the
rms of the difference between the measured and input $b$.  The middle
panels show the median of the difference between the measured and
input $b$ (an indication of systematic errors in the fits).  The
different curves are for different photometric precision $\sigma =
0.015$ (pentagons), 0.01 (asterisks), 0.005 (open triangles), 0.0025
(solid triangles), and time sampling is shown in the $x$-axis.  Notice
the systematic under-estimate of $b$, especially evident for the $b
=0.7$ models with large $\sigma$ and $\delta t$.  The bottom panels
show normalized histograms of the of fit values for $b$ for the
different combinations of $\delta t$ (increasing from left to right)
and $\sigma$ (increasing from bottom to top).  The dotted line in each
sub-panel indicates the input value for $b$.  The figure shows that
very high time sampling and high photometric precision are needed for
a reasonably accurate fit ($\lesssim 10$--$20$\% errors). When either
the time sampling or photometric precision are very low, the value of
$b$ is consistently underestimated by the $\chi^2$ fits (see text and
Figure \ref{fig:bvstf}). The panels in this figure can be used to estimate
errors in parameters from different models (for the same $b$) by
considering that changing $t_T$ causes a linear change in time
sampling and changing $\Delta F$ causes a linear change in error in
photometric precision.  Note the trade off in $\sigma$ vs. $\delta t$,
where combinations with the same $\sigma^2 \times \delta t$ have
nearly identical error distributions. }
\label{fig:berror}
\end{figure}

\begin{figure}
\plotone{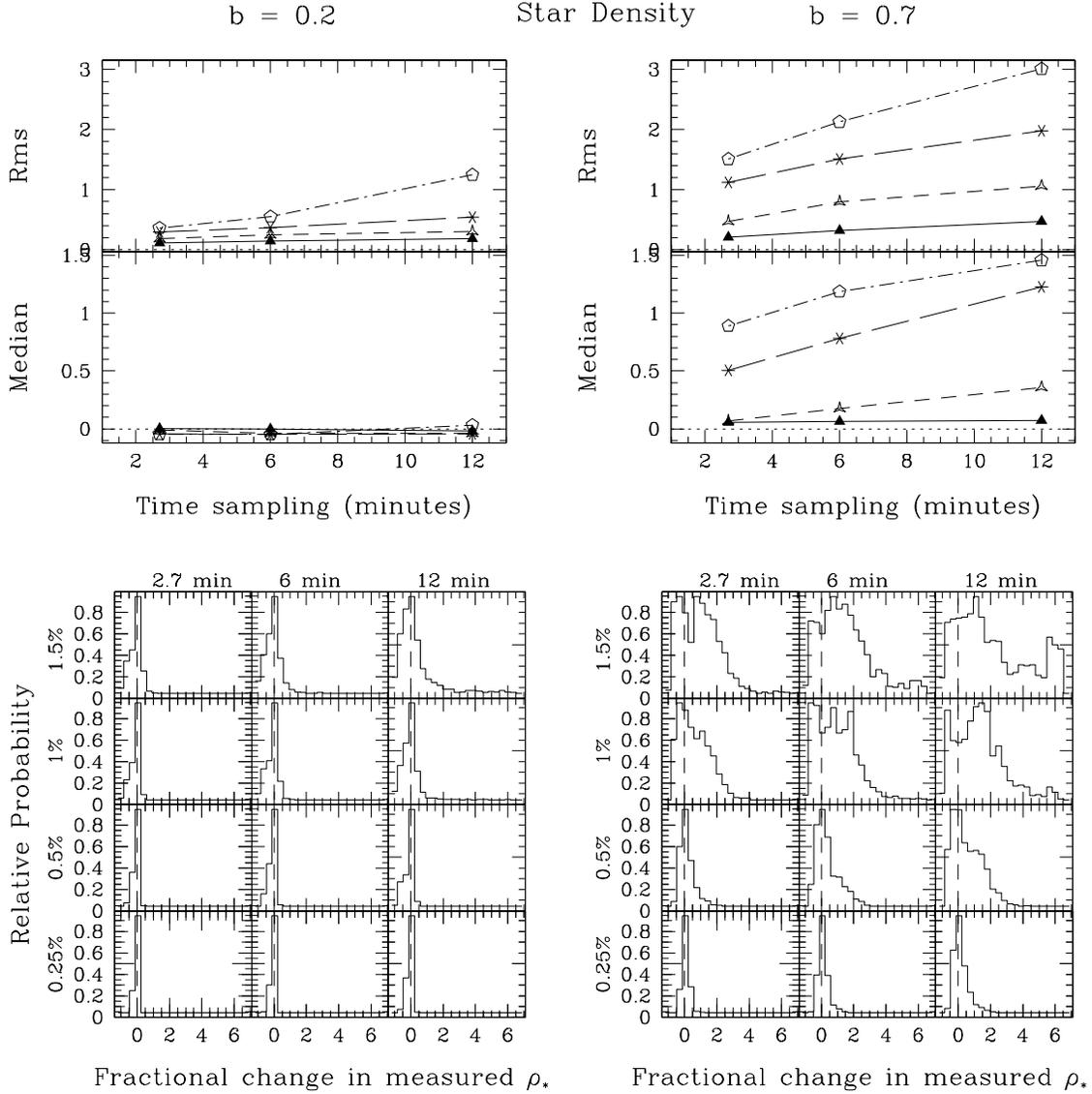}
\caption{The fractional errors in the derived stellar density. See
caption of Figure~\ref{fig:berror} for details.  Notice that for low
photometric precision and time sampling the density is significantly
over-estimated for the $b = 0.7$ case.  This is a consequence of the
under-estimate of $b$ (see Figure~\ref{fig:berror}).}
\label{fig:rhoerror}
\end{figure}

\begin{figure}
\plotone{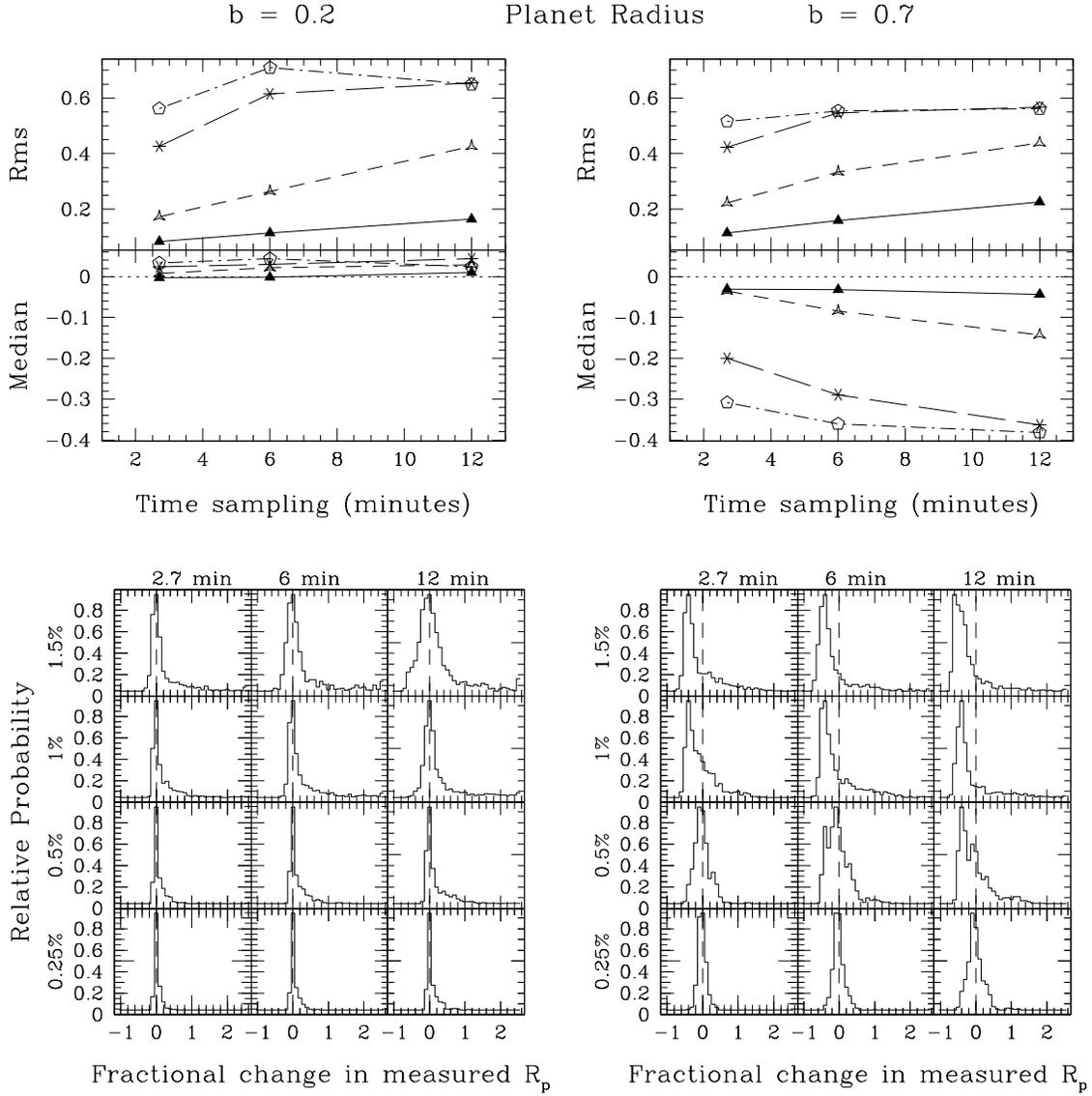}
\caption{The fractional errors in the derived planet radius. See
caption of Figure~\ref{fig:berror} for details.  Notice that for low
photometric precision and time sampling the planet radius is
consistently under-estimated for the $b = 0.7$ case.  This is a
consequence of the under-estimate of $b$ (see
Figure~\ref{fig:berror}).  This under-estimate will result in some M
dwarfs being classified as planet-sized when using data with low
photometric precision and time sampling.}
\label{fig:rperror}
\end{figure}

Figures~\ref{fig:berror}, \ref{fig:rhoerror} and
\ref{fig:rperror} show the errors in $b$ and the fractional errors
in $\rhostar$ and $\rp$ for the simulations on the specific transit
model described above.  Results for $b = 0.2 $ are shown on the left
panels, and results for $b = 0.7$ are shown on the right panels.  The
top and middle panels respectively show the rms and the median of the
difference between the fit results and the input parameter (e.g., the
rms and median of \{$b_i - b_{model}$\}, where the $b_i$ are the fit
results from each of the 1000 simulated noisy light curves, and
$b_{model}$ is the actual value used to create those light curves).
The bottom panels show histograms of the fit results for each
combination of photometric precision and time sampling.  The vertical
dotted lines indicate the correct value for the parameter in question
(0.2 or 0.7 for $b$, and 0 in the case of fractional deviation for
$\rhostar$ and $\rp$).  From the median deviation plots and the
histograms themselves, notice the severe systematic underestimate of
$b$ and the resulting over-estimate of $\rhostar$ and $\rp$ for cases
with $\sigma \gtrsim 0.005$~mag and $b = 0.7$.  The rms fractional
errors in $\rhostar$ are 10\% for $b=0.2$ and 20\% for $b=0.7$ for a
photometric precision of 0.0025 and a time sampling of 2.7 minutes.
For this photometric precision and time sampling, the errors in $\rp$
are less than 10\% (neglecting uncertainty in the stellar mass-radius
relation). The errors in $\rp$ are $\lesssim 40$\% for time samplings
$\lesssim 6$ minutes and photometric precision $\lesssim 0.005$.
These errors are quoted for the model used here, and as described
above errors for models with different parameters (with the exception
of changes in $b$) can be derived by scaling $\delta t$ and $\sigma$
together with $t_T$ and $\Delta F$.  Note that folded transits from multiple
low-time-sampled transit light curves can be used for higher effective
time sampling and are useful as long as the photometric precision is
high enough.  A main point of these error simulations is that the
transit shape must be well defined by high photometric precision and
high time sampling, because most star-planet parameters depend on
transit shape. Specifically a limiting factor is time sampling of
ingress and egress and their start and end times to determine transit
shape.

The errors for different combinations of photometric precison
($\sigma$) and time sampling ($\delta t$) are related because the
total S/N per transit is based on both $\sigma$ and $\delta t$.  In
particular the total number of photons per transit goes roughly as:
\begin{equation} 
N_{photons \hspace{0.03in} per \hspace{0.03in}
transit} \sim \left( \frac{t_T}{\delta t}\right)
\left(\frac{1}{\sigma^2}\right).  
\end{equation} 
It is interesting to note that our simulations show that the same
error distribution will result from $\delta t$ and $\sigma$
combinations that give the same total number of photons per transit.
E.g., compare the errors for a given $\delta t$ and $\sigma$ with
those for $\delta t \times 4$ and $\sigma/2$. This can be seen in
Figure~\ref{fig:rperror} where the same errors and error distribution
results for $\delta t = 12$ mins, $\sigma = 0.0025$~mag and for $\delta t
= 2.7$ mins, $\sigma = 0.005$~mag.  Taking this into account, an optimal
strategy of time sampling and exposure time can be chosen for a given
mirror size and detector read time.  Note that in practice, $\delta t
\ll t_T$\ is required in order to reach reasonable errors in the
derived parameters.

\section{Limb Darkening}
\label{sec-ld} 
\begin{figure}
\plotfiddle{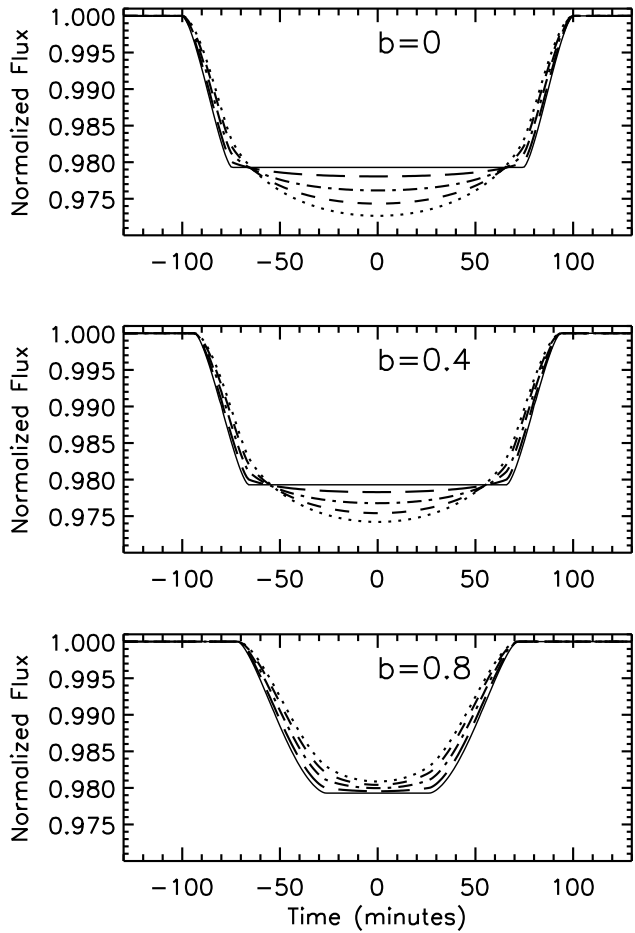}{5in}{0}{95}{95}{-350}{-300}
\caption{Solar limb darkening dependence of a planet transit light
curve.  In these theoretical light curves the planet has $\rp=1.4R_J$
and $a=0.05$~AU and the star has $\rstar=R_{\odot}$ and
$\mstar=M_{\odot}$.  The solid curve shows a transit light curve with
limb darkening neglected. The other planet transit light curves have
solar limb darkening at wavelengths (in $\mu$m): 3, 0.8, 0.55, 0.45.
From top to bottom the panels show transits with different impact
parameters $b$, which correspond to inclinations $\cos i = b
\rstar/a$.  Although the transit depth changes at different
wavelengths, the ingress and egress slope do not change significantly;
the different slopes are generally equivalent within typical
observational errors.  The ingress and egress slope mainly depend on
the time it takes the planet to cross the stellar limb.}
\label{fig:ld}
\end{figure}

The transit light curve equations, their solution,
and applications described in this paper assume no limb darkening.
For the framework used in this paper, $t_F$ (see
Figure~\ref{fig:schema}) must be measureable from the light
curve---this is only possibly at wavelengths where limb darkening is
weak. Figure~\ref{fig:ld} shows that $t_F$ and $t_T$ are
distinguishable for measurements at $I$-band. At bluer band passes
limb darkening becomes significant and it is not clear when the
ingress/egress ends and begins and when the planet is fully
superimposed on the star.

It is possible to incorporate limb darkening into the mathematical
description of a planet transit light curve by parameterizing the
transit shape by the slope of the ingress/egress instead of by
$t_F$ and $t_T$. The slope of ingress/egress is given by the time
derivative of the transit area equation, including a
parameterization for limb darkening. Figure~\ref{fig:ld} shows
that the slope of ingress/egress is similar for a planet transit
in different colors; the different slopes are generally equivalent
within typical observational errors. This is because, for $\rp \ll
\rstar$, the ingress/egress slope is mainly due to the time it
takes the planet to cross the stellar limb. In addition, the depth
of the transit is affected by limb darkening. As a
consequence of these two additions the equations that describe
the transit light curve take a cumbersome
form and there is no longer a simple analytical solution. Of
course a model fit with limb darkening can still be used to
determine the parameters of the system. However, for a quick
parameter estimate and the applications described in
\S\ref{sec-applications} an analytic solution is simple and
practical; $I$-band or redder observations can be used to
circumvent the complications of limb darkening.

\section{Applications}
\label{sec-applications}
 The existence of a unique solution to a
planet transit light curve has several applications. The first
application is an analytic solution for planet-star parameters by
using the equations in \S\ref{sec-fivesoln} with the values from
\S\ref{sec-foursoln} or \S\ref{sec-simpsoln}. This analytical
solution is useful for a quick parameter estimate. In this section
we use the simplified analytic solution (\S\ref{sec-approx}) to
explore the usefulness of four additional applications.

\subsection{Measuring Stellar Density Directly from the Transit Light Curve}
\begin{figure}
\plotone{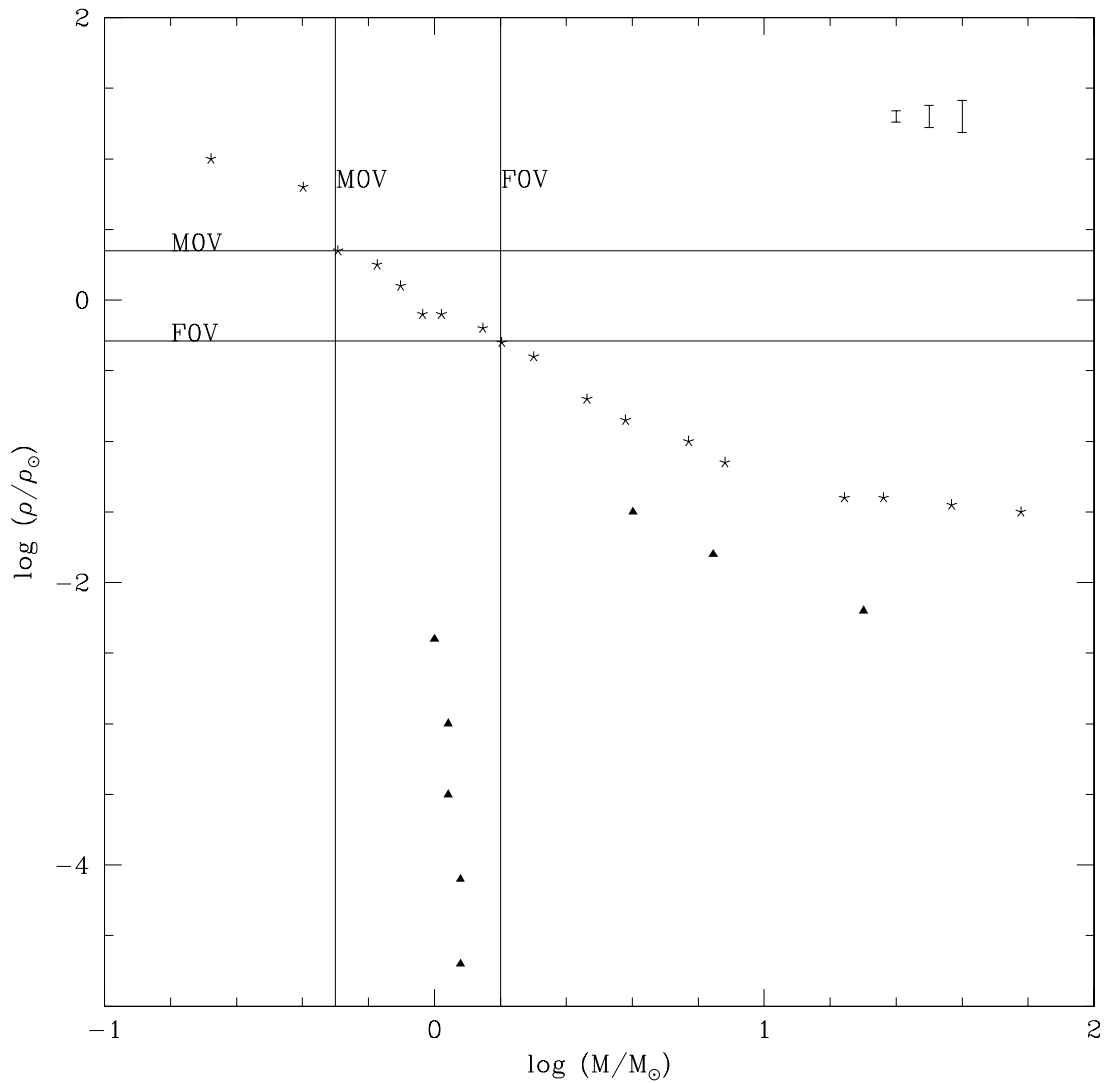} 
\caption{Stellar density $\rho_*$ vs. star mass $M_*$. The asterisks
show main sequence stars and the triangles show giant stars (Cox 2000).
The unique solution of $\rhostar$ from the planet transit light curves
will give a position on the $y$-axis. The box M0V to F0V shows the
main sequence stars which are most appropriate for finding transiting
planets.  These stars have a very different density from giant stars,
and are therefore easily identifiable by their density alone.
Although stars that have slightly evolved off of the main sequence
will also populate the M0V/F0V box, their radii can still be estimated
to $\lesssim 25\%$ accuracy (see text).   The error bars in the
upper right corner are for fractional errors in $\rhostar$ of 0.1,
0.2, and 0.3.  See Figure~\ref{fig:rhoerror} for the errors in $\rhostar$.
} 
\label{fig:rhovsmass}
\end{figure}

The density $\rho_* = \mstar \rstar^{-3}$, as described analytically
in equations~(\ref{eq:rhoorig}) and (\ref{eq:rhoapprox}), is directly
measurable from the light curve observables ($\Delta F$, $t_T$, $t_F$,
and $P$), together with Kepler's Third Law. A measured density makes
it possible to immediately distinguish between a main sequence star
and a giant star from the transit light curve alone (i.e., without
color data or spectra). Knowing the stellar size to first order
immediately is useful for an estimate of $\rp$ from the depth of the
transit (equation~(\ref{eq:rporig})). For example, eclipses of $\Delta
F = 1$\% for a giant star would be caused by a stellar companion, not
a planetary companion. Figure~\ref{fig:rhovsmass} shows the density as
a function of spectral type. A density measurement would give a
position on the $y$-axis only. As shown by the box in
Figure~\ref{fig:rhovsmass}, F0V to M0V stars occupy a unique region of
stellar density parameter space. Hence a density measurement can tell
us that a given star is in the vicinity of main sequence stars. The
main sequence stellar spectral type can be estimated for a given
$\rhostar$ by using the mass-radius relation for main sequence stars.

The stellar type (and hence $\rstar$ and $\rp$) may not be identified
with 100\% certainty from $\rhostar$ alone due to confusion with stars
slightly evolved off of the main sequence; the unique density
derivation clearly does not specify the actual stellar mass and
radius. As main sequence stars begin to leave the main sequence they
will evolve with constant mass to lower density as their radius
increases (on vertical downward lines in
Figure~\ref{fig:rhovsmass}). These slightly evolved stars will fill in
the box in Figure~\ref{fig:rhovsmass} to the lower left of the main
sequence star diagonal line.  Nevertheless, for a given $\rhostar$ the
corresponding $\rstar$ of a main-sequence star will always be the
upper limit of $\rstar$, whether or not the star is slightly evolved.
Thus, we can always determine an upper limit to $\rstar$---and hence
to $\rp$---from $\rhostar$ alone.  Furthermore, a lower limit to
$\rstar$ for a given $\rhostar$ can be derived with the consideration
that stars with $\mstar < 0.8 \msun$ do not leave the main sequence in
a Hubble time.  Note that for densities corresponding F0V to M0V
stars, the largest error in the radius estimation will occur for an
evolved $0.8 \msun$ star with the density of an F0V star; in this
extreme scenario, the radius over-estimate of the evolved star is only
25\%.  Beyond using the light curve alone, a derived surface gravity
from a spectrum (measured to $\lesssim 20$\% precision for non-metal
poor stars near the main sequence (Allende Prieto et al.\ 1999)) can
be used with $\rhostar$ derived from the transit light curve to
determine $\mstar$ and $\rstar$. Note that highly evolved stars will
never be confused with main sequence or slightly evolved stars, since
$\rhostar$ will be significantly lower for highly evolved stars.

The errors for $\rhostar$ are described in \S\ref{sec-errors} and
plotted in Figure~\ref{fig:rhoerror}. Figure~\ref{fig:rhovsmass}
shows a plot of density vs. main sequence spectral type, together
with reference error bars. With an error in $t_F$ and $t_T$ of $<$ 5
minutes, and errors in $\Delta F <1$\%, the stellar density can be
constrained to 10\%--20\% depending on the transit shape (see
\S\ref{sec-errors}). A consequence of the unique solution
of $\rhostar$ is that transit fitting codes which find equally
good fits for different combinations of $\mstar$ and $\rstar$
actually find {\it the same $\rhostar$} for these best fits.

\subsection{Contamination from Blended Stars: Transit Light Curve Alone}
\label{sec-blendslc} 
\begin{figure}
\plotfiddle{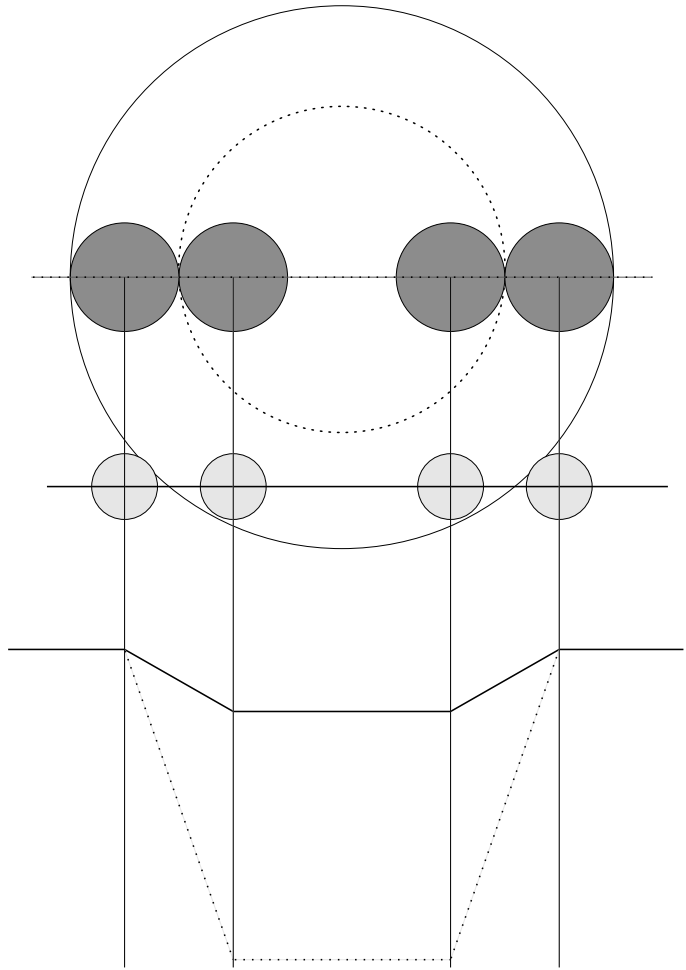}{5in}{0}{65}{65}{-120}{50}
\caption{The effect of a blended star on a transit light curve.  A
schematic light curve (dotted line at the bottom) shows a relatively
deep eclipse due to an eclipsing binary star system in which a small
star centrally crosses the smaller primary star (dotted circle).  In
the presence of additional light from a third blended star (not
shown), this deep eclipse will appear shallower, as shown by the solid
line transit curve.  This shallower solid-line transit looks the same
as a planet transit crossing a larger star (solid line circle) with
high impact parameter.  Thus, there is no longer a unique solution to
a transiting system in the presence of a blend.  Note that any such
blended transit will never look box-shaped, since the long ingress and
egress are caused by the relatively long time it takes for the
secondary star to cross the limb of the primary star.}
\label{fig:blenddef}
\end{figure}

The five-parameter solution to a planet
transit light curve with two or more transits is {not unique} if
there is additional light from a blended star (or indeed if any of
the assumptions in \S\ref{sec-assumptions} are not fulfilled). A
planet transit light curve can be mimicked by a stellar binary
eclipsing system with contaminating light from a third, fully
blended star. The additional light from the blended star causes an
otherwise deep eclipse to appear shallow, and hence to mimic a planet
transit light curve, as shown in Figure~\ref{fig:blenddef}. This
confusion is possible only when the eclipsing system has a
flat-bottomed light curve, meaning that the eclipsing companion is
fully superimposed on its primary star. Hereafter we call the
eclipsing system plus the third fully blended star a ``blend''.

\subsubsection{Star-planet Parameters and Blends}
The true, or actual, star-planet parameters will be confused by
the contamination of blended light from an additional star. For
selecting planet candidates for radial velocity follow-up it is
useful to consider how some of the derived star and planet parameters
change in the presence of blended light. In the following
discussion we compare star and planet parameters naively derived
from a transit light curve assuming no blend against those parameters
derived by assuming the maximum possible amount of blended light.

\begin{figure}
\plotfiddle{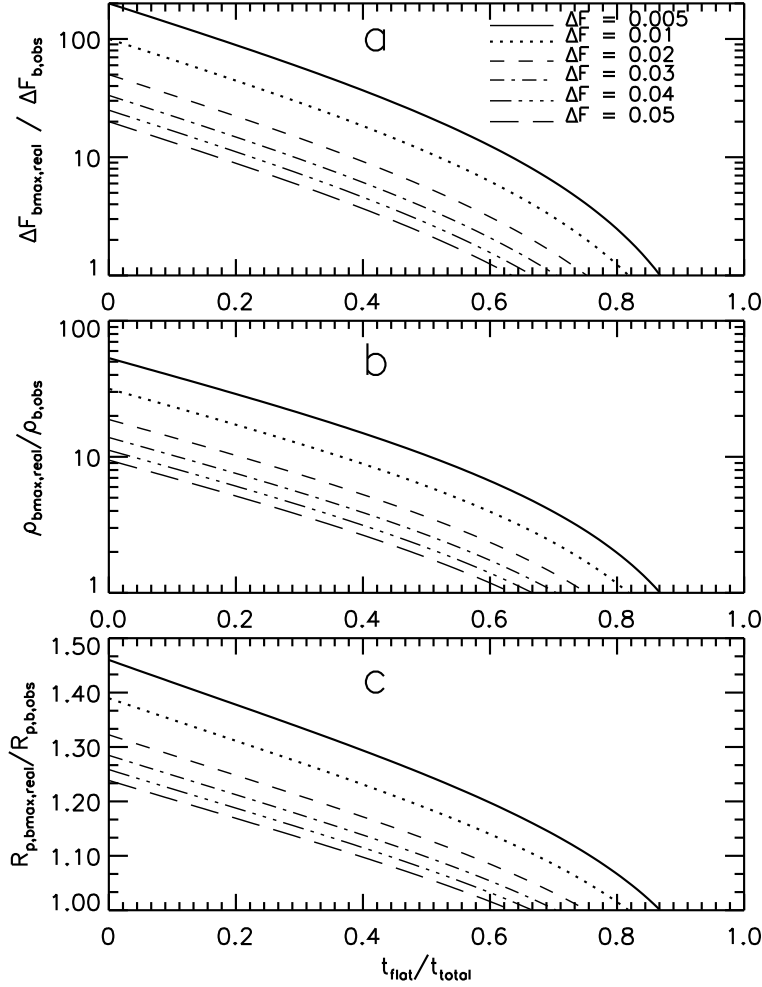}{5in}{0}{95}{95}{-250}{-300}
\caption{The maximum effect of a third fully blended star on the
naively derived eclipsing system's parameters $\Delta F$, $\rho_*$ and
$\rp$. Panel a: The ratio between the maximum real eclipse depth in
the presence of a blend $\Delta F_{b, real, max}$ and the observed
eclipse depth $\Delta F_{b,obs}$ as a function of transit shape
parameterized by $t_F/t_T$ (see
equation~(\ref{eq:maxblenddepth})). For a small transit depth (e.g.,
$\Delta F = 0.005$) and a small $t_F/t_T$ (i.e., a nearly triangular
transit shape) a large amount of extra light can be hidden in the
transit light curve.  Panel b: The ratio between the maximum real
stellar density $\rho_{b, real, max}$ and the naively-derived density
$\rho_{b,obs}$ in the case of a blend.  This ratio depends on both
$t_F/t_T$ and $\Delta F$ (equations~(\ref{eq:blendlimit}) and
(\ref{eq:maxblenddepth})).  Note that the real stellar density is
always higher than the ``observed'' stellar density naively derived
from a light curve ignoring the blend.  Panel c: The ratio between the
maximum real companion radius, $R_{p,b, real, max}$ in the presence of
a blend and the naively-derived radius $R_{p,b,obs}$, assuming the
mass-radius relation for main sequence stars (see
equation~(\ref{eq:rpb})). Note that in the case of a blend, the real
eclipse depth, real stellar density, and and real companion radius are
always larger than the naively-derived observed values.}
\label{fig:blendrp}
\label{fig:rhovssp}
\end{figure}

The maximum amount of blended light for a given $\Delta F$ is computed
by considering the largest possible companion consistent with a given
ingress/egress duration.  This is equivalent to assuming a central
transit (i.e., $b = 0$), and computing the maximum $\Delta F$ possible
for that given ingress/egress (i.e., $t_F/t_T$). Specifically, using $b=0$ in
equation~(\ref{eq:shapeapprox}),
\begin{equation}
\label{eq:maxblenddepth}
\Delta F_{b,real} \leq \Delta F_{b, real, max} = \frac{\left(1 -
\frac{t_F}{t_T} \right)^2}{\left(1 + \frac{t_F}{t_T} \right)^2}.
\end{equation}
Here the subscript $b$ refers to a blend; the subscript $obs$ refers
to the eclipsing system parameters derived ignoring the presence of
the third fully blended star whereas the subscript $real$ refers to
the actual parameters of the eclipsing binary.  The subscript $max$
refers to the quantities for the case of maximum possible blend
(defined for a given $t_F/t_T$). $\Delta F_{b,obs}$ is the eclipse
depth as taken from the observed light curve, and $\Delta F_{b,real}$
is the actual eclipse depth (i.e., in the absence of the blended star).
The ratio $\Delta F_{b, real, max} / \Delta F_{b, obs} $ is shown in
Figure~\ref{fig:blendrp}a.

It is useful to consider the maximum difference
between the density derived from a transit light curve in the
presence of a blend ($\rho_{b,obs}$) and the actual density of the
eclipsed star ($\rho_{b,real}$). We do this by
using the simplified equation for $\rhostar$
equation~(\ref{eq:rhoapprox}), to compute the ratio
$\rho_{b,obs}/\rho_{b, real, max}$:
\begin{equation}
\label{eq:blendlimit}
 1 \leq  \frac{\rho_{b,real}}{\rho_{b,obs}}
\leq \frac{\rho_{b,real,max}}{\rho_{b,obs}}= \left[ \frac{\Delta
F_{b,real,max}}{\Delta F_{b,obs}} \right ]^{3/4}.
\end{equation}
The stellar density is always underestimated from the light curve if a
blend is present and ignored. A comparison of $\rho_{b,real,max}$ and
$\rho_{b,obs}$ is shown in Figure~\ref{fig:rhovssp}b and is used in an
application in subsection \ref{sec-blendslcsp}.

We can explore how the planet radius is affected by blends by
considering how $\rhostar$ is affected by blends
(equation~(\ref{eq:blendlimit})) together with the simplified solution for
planet radius, equation~(\ref{eq:rporig}). The actual radius of
the eclipsing companion, $R_{p,b,real}$, is related to the radius
naively derived from the transit light curve ignoring the presence
of a blend, $R_{p,b,obs}$, by
\begin{equation}
\label{eq:rpb}
\frac{R_{p,b,real}}{R_{p,b,obs}}
= \left[ \frac{\Delta F_{b,real}}{\Delta F_{b,obs}} \right ]
^{\frac{3x-2}{12x-4}}
\leq \left[ \frac{\Delta F_{b,real,max}}{\Delta F_{b,obs}} \right
]^{\frac{3x-2}{12x-4}}, 
\end{equation}
where $x$ is the exponent in the stellar mass-radius relation
(equation~(\ref{eq:MRorig})).  For main sequence stars, $x \approx
0.8$ (Cox 2000) and the exponent in equation~(\ref{eq:rpb}) becomes
0.071, so $R_{p,b,real}/R_{p,b,obs}$ is mildly dependent on the amount
of blended light present.  The fact that there
is a maximum value for $\Delta F_{b,real}$ together with
equation~(\ref{eq:rpb}) implies that $R_{p,b,real}$ also has a maximum
for a given $t_F/t_T$ and $\Delta F$.  Figure~\ref{fig:blendrp}c shows
$R_{p,b,real,max}/R_{p,b,obs}$ as a function of transit shape
parameterized by $t_F/t_T$. In the presence of a blend, the actual
radius of the eclipsing companion is always larger than the radius
derived from the eclipsing light curve.  Thus this figure shows the
importance of ruling out blends---{\it the eclipsing companion in a
blended case could be larger than expected for a close-in planet ($\rp
\sim R_J$) by 20--50\%}.

\subsubsection{Probability of Blends and Blend Limits}
Transits with short ingress and egress times (box-shaped transits with
large $t_F/t_T$) are least likely to be affected by blended stars and
can be the best planet transit candidates for follow-up radial
velocity mass measurements. There are two reasons for this.  First,
less blended light can be hidden in a shallow transit with large
$t_F/t_T$ than one with small $t_F/t_T$. This is seen directly from
the maximum value of $\Delta F_{b,real,max}$ given in
equation~(\ref{eq:maxblenddepth}) and shown in Figure~\ref{fig:blendrp}a.

\begin{figure}
\plotone{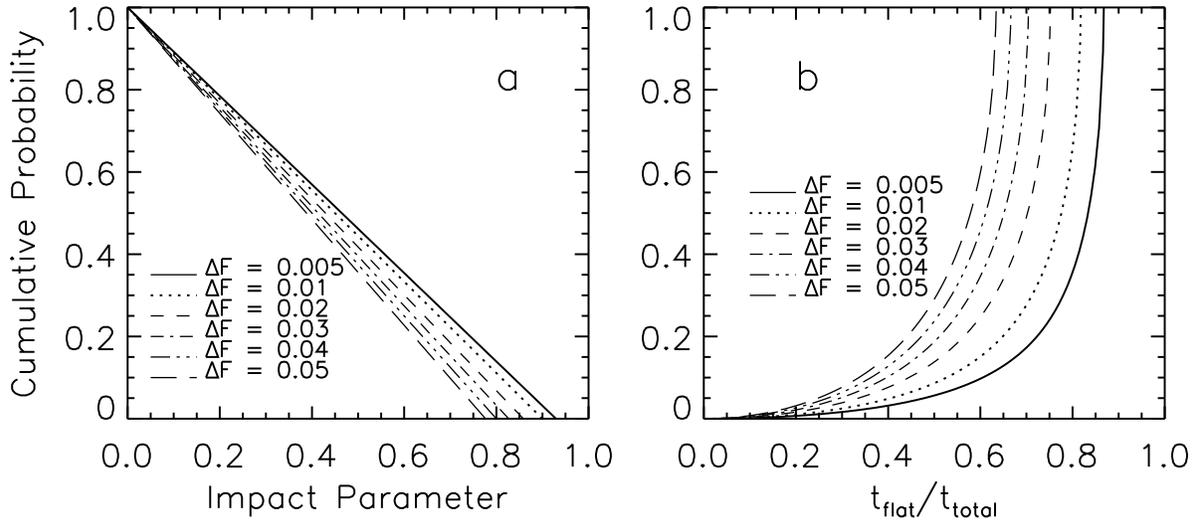} 
\caption{Panel a: the cumulative probability ($P_c$) that, for a
random orientation of orbital inclinations, a transit has impact
parameter $b$ which is $larger$ than a specified impact parameter
$b_x$ (shown on the $x$-axis).  This cumulative probability is just
due to geometry.  Panel b: the same cumulative probability as a
function of transit shape, as parameterized by $t_F/t_T$.  $P_c$ can be
described by both $b$ and $t_F/t_T$ because there is a one-to-one
correspondence between them for a given transit depth, as shown in
Figure~\ref{fig:bvstf}. The $P_c$ shows that ``box-like'' transit
shapes (i.e., high $t_F/t_T$) are much more common than transits with
long ingress and egress (i.e., transits with small $t_F/t_T$).  If in
a transit survey large fraction of transits of a given $\Delta F$ 
are found to have small $t_F/t_T$, this is an indication that
most of these systems are likely blended (see Figure~\ref{fig:blenddef}).}
\label{fig:probplanet}
\end{figure}

The second reason why box-shaped transits are least likely to be
affected by blended stars is because of the geometric probability
for different transit shapes given random orientations of orbital
inclinations.
First it is useful to recognize that the unique solution to the
planet-star parameters is in part possible because for a 
given $\Delta F$ there is a
one-to-one correspondence between the impact parameter $b$ and the
transit shape as parameterized by $t_F/t_T$. This $b$ vs.
$t_F/t_T$ relation is non linear, as shown in
Figure~\ref{fig:bvstf}. For a random orientation of orbital
inclinations, the cumulative probability $P_c$ that an unblended
transit will have impact parameter $b$ {\it smaller} than some given value $b_x$
is proportional to $b_x$;  the maximum value of $b$ for a transit is $
1 - \rp / \rstar = 1 - \sqdepth$, where $\sqdepth$ is the value for an
unblended transit.  Thus, the cumulative
probability that an unblended transit will have $b$ {\it larger} than
$b_x$ is described by
\begin{equation}\label{eq:probplanet}
\label{eq:Pc} P_c (b \leq b_x) = \frac{1 - \sqrt{\Delta F} -
b_x}{1 - \sqrt{\Delta F}},
\end{equation}
and is shown in Figure~\ref{fig:probplanet}a.  Because of the
one-to-one correspondence between $b$ and transit shape, we can also
express the cumulative probability in terms of $t_F/t_T$ (i.e.,
transit shape), shown in Figure~\ref{fig:probplanet}b. The cumulative
probability in equation~(\ref{eq:Pc}) and Figure~\ref{fig:probplanet}
can be used to show that for an ensemble of transits, there should be
more ``box-shaped'' transits (large $t_f/t_T$) than transits with very long
ingress/egress times (small $t_F/t_T$).

The S/N that must be reached to be able to completely rule out a
possible blend can be determined from the maximum real eclipse depth $\Delta
F_{b,real,max}$.  The light from the eclipsing primary star
is always less than or equal to the total amount of light from the
blended system, and has a minimum value given by the maximum blend:
\begin{equation}
1 \ge \frac{F_{primary}}{F_{b,total}} = \frac {\Delta F_{b,obs}}
{\Delta F_{b,real}} \ge \frac {\Delta F_{b,obs}}{\Delta F_{b, real,
max}}.
\label{eq:primtot}
\end{equation}
Thus, the quantity ${\Delta F_{b,obs}}/{\Delta F_{b, real, max}}$ gives
a lower limit to the fraction of the observed light which comes from
the primary star in the eclipsing system.  This can be used to
establish the data quality necessary to find or rule out a blend in a
given case.  The light from the eclipsing primary in the blend could
be detected, for example, by cross-correlating spectra of the blend
with a suite of stellar templates to see if two cross-correlation
peaks are present, or by resolving the blend with high spatial
resolution imaging (e.g., from space or adaptive optics).

To summarize this subsection: \\
$\bullet$ box-shaped transits are the best planet
candidates for radial velocity mass follow-up because (1) the least
amount of blended light can be hidden in box-shaped transits and (2)
for an ensemble of randomly oriented orbital inclinations there should
be many more box-shaped transits with large $t_F/t_T$ than transits
with small $t_F/t_T$. \\ 
$\bullet$  In the presence of a blended star, the real
planet radius is always larger than the naively derived planet radius
ignoring the blend.\\
$\bullet$ The maximum brightness of a possible contaminating
star can be derived from the transit light curve.

\subsection{Contamination from Blended Stars: Transit Light Curve {\it and} a Known Spectral Type}
\label{sec-blendslcsp} Knowing the spectral type independently of
the parameters derived from the transit light curve can be
extremely useful in ruling out blends. With $\mstar$ and $\rstar$
determined from a spectral type, the system of
equations~(\ref{eq:depthorig})--(\ref{eq:Keplerorig}) is
overconstrained.  If the stellar density $\rho_{b,obs}$ as derived
from the light curve from equation~(\ref{eq:rhoorig}) is very
different from the density $\rho_{spect}$ of the observed spectral
type then something is amiss with the light curve assumptions in
\S\ref{sec-assumptions}. A possibility is the case of a blended
star discussed above. Thus comparing $\rho_{obs}$ with $\rho_{spect}$
is extremely useful for detecting the presence of a blend.

For example the density can be overestimated a factor of up to
10 for $t_F/t_T = 0.4$ and $\Delta F=0.1$.  The maximum density ratios
on Figure~\ref{fig:rhovssp}b can be compared directly with the actual
stellar main sequence densities shown in
Figure~\ref{fig:rhovsmass}. Furthermore, by comparing $\rho_{obs}$ and
$\rho_{spect}$ we can detect the presence of a blend to an upper limit
determined by the fractional errors in the stellar density, $\frac {\delta \rhostar}{\rhostar} \simeq
\sqrt{\left ( \frac {\delta \rho_{obs}}{\rho_{obs}}\right ) ^2 + \left ( \frac{\delta
\rho_{spect}}{\rho_{spect}}\right ) ^2}$. From
equations~(\ref{eq:primtot}) and (\ref{eq:blendlimit}) we can see that
it will be possible to detect blends with 
$ \frac{F_{primary}}{F_{b,total}} = \frac {\Delta F_{b,obs}}
{\Delta F_{b,real}} \gtrsim (1 - \frac{\delta
\rhostar }{ \rhostar})^{4/3}$.   Note that the error in $\rho_{spect}$
must account for the fact that the spectrum may be dominated by the 
blended star, rather than by the primary star in the eclipsing system.

\subsection{Blends and Statistics}
\label{sec-statistics} Blended stars may be a common contaminant
in transit searches (D. Latham, private communication 2001;
Mall\'en-Ornelas et al., in preparation) either due to very wide
binaries or, less likely in an uncrowded field, due to a chance
alignment of a foreground or a background field star. In the presence
of a third fully blended star some planet transits will be ``washed
out'' and not detectable. Hence it is important to know the blend
frequency to determine the frequency of transiting planets. Even
without a measured stellar spectral type, the frequency of blends
contaminating the sample of planet candidates can be determined
statistically by considering all of the flat-bottomed eclipses of
various depths in the following way. Assuming the orbital inclination
distribution is uniform, for a given $\Delta F$ light curves should
have a certain distribution of $t_F/t_T$ (see the probability
discussion in
\S\ref{sec-blendslc}). Deviations from each expected distribution
can be explained by eclipses being made more shallow than they should
be by contaminating blended stars. The fraction of blended stars will
also allow a statistical estimate of how many transits may be missed
due to blends.  

\subsection{Period Derivation from Single-Transit Light Curves and a Known Spectral Type}
The period can be estimated for a light curve with only one transit if
$M_*$ and $R_*$ are known from a measured spectral type. Again, this
is because the system of
equations~(\ref{eq:depthorig})--(\ref{eq:Keplerorig}) that describe
the transit light curve are overconstrained when $M_*$ and $R_*$ are
known. Considering that $M_*$ and $R_*$ are known, and that the impact
parameter is derived from equation~(\ref{eq:borig}), the period can be
solved for from the following function of P:
\begin{equation}
\frac{4 \pi^2}{P^2 G}
\left[ \frac{ (1 + \sqrt{\Delta
F})^2 - b^2 (1 - \sin^2 \frac{t_{T} \pi}{P})} { \sin^2
\frac{t_{T} \pi}{P}} \right]^{3/2} =
\frac{\mstar}{\rstar^3},
\end{equation}
where the right hand side can be determined from the stellar
spectral type.
Under the approximation $t_T \pi/P \ll 1$ the period is simply given by
\begin{equation}
\label{eq:papprox}
 P = \frac{\mstar}{\rstar^3} \frac{G \pi}{32}
 \frac{\left(t_T^2 - t_F^2\right)^{3/2}}{\Delta F^{3/4}}.
\end{equation}
For $P$ in days, the first term on the right hand side is $\frac{G
\pi}{32} = 288.73$. The error for this $P$ estimate is not easily
derived because the errors in $t_F$ and $t_T$ are highly
non-Gaussian and are often correlated. The fractional error in the
period, $\delta P / P$, can be estimated for small errors in time
sampling and photometric precision by noting that
equation~(\ref{eq:papprox}) is derived from the simplified
equation for $\rhostar$ (equation~(\ref{eq:rhoapprox})), and that
$\mstar$ and $\rstar$ would be known from the spectral type.
$\delta P / P  = \sqrt{ \left ( \frac {\delta
\rho_{obs}}{\rho_{obs}}\right ) ^2 + \left ( \frac {\delta
\rho_{spect}}{\rho_{spect}}\right ) ^ 2}$,
where $\delta \rho_{obs}/\rho_{obs}$ is the fractional error in $\rhostar$
derived from the transit light curve and $\delta \rho_{spect}/\rho_{spect}$
is the fractional error in $\rhostar$ derived from the spectral
type. Thus, following the discussion in \S\ref{sec-errors} and
assuming $\delta \rho_{spect} / \rho_{spect} \lesssim 0.1$, the period can be
determined to $\sim$ 15--20\% (depending on transit shape) for
$\delta t < 5$ minutes and $\sigma \sim 0.0025$~mag.

An upper limit
to the period can be estimated even without knowing $t_F$,
\begin{equation}
P \leq \frac{G \pi}{32} \frac{\mstar}{\rstar^3}
 \frac{t_T^2}{\Delta F^{3/4}}.
\end{equation}
This upper limit is valid even in the case
of a blend where $\Delta F_{b,real}$ is
larger than the observed $\Delta F$.

\section{Summary}
We have presented the equations that describe a light curve with two
or more transits and have presented the unique solution for the impact
parameter $b$, the ratio of the orbital distance to stellar radius
$a/\rstar$, and stellar density $\rhostar$.  Furthermore, with the
stellar mass-radius relation we can uniquely derive the five
parameters $\mstar$, $\rstar$, $i$, $a$, $\rp$. This unique solution
is only possible under the assumptions listed in \S2, most importantly
that the light curve is from a single star (i.e.  not two or more
fully blended stars), that the planet is dark and is in a circular
orbit, and the light curve transits have flat bottoms (which can be
obtained at red or longer wavelengths).

We have found: \\
$\bullet$ A simple analytical solution that can be used to quickly
estimate the planet-star parameters, most importantly $\rhostar$ and $\rp$; \\
$\bullet$ The stellar density can be uniquely determined from the
transit light curve alone. Fitting codes that solve for star-planet parameters
will find a number of best fits for different combinations of
$\mstar$ and $\rstar$---these best fits will have the same stellar
density. \\
$\bullet$ For noisy data, the impact parameter $b$, $\rstar$ and
$\rp$ are underestimated and $\rhostar$ is overestimated due to a
non-linear one-to-one correspondence for a given $\Delta F$
between $b$ and transit shape (as parameterized by $t_F/t_T$).

The existence of the unique solution  for the above parameters
allows several interesting applications, including: \\
$\bullet$ The stellar radius and the planet radius can be estimated
based on $\rhostar$;\\
$\bullet$ The likelihood that a shallow transit is due
to contamination from a fully blended star can be estimated---with
box-shaped transits being the least likely to be contaminated by blends; \\
$\bullet$ A comparison of $\rhostar$ as determined from the transit light curve
with $\rhostar$ from a spectral type can help identify
blended stars and hence indicate that shallow eclipses are not
due to planet transits; \\
$\bullet$ In the presence of a blend the actual eclipsing
companion radius is always larger than the radius derived from the
light curve; \\
$\bullet$ The period from a single transit event can be estimated
with a known
spectral type. \\

For most of these applications time sampling of $\delta t < 5$ minutes
and photometric precision of $\sigma < 0.005$~mag are needed (or more
generally $\sigma^2 \times \delta t \lesssim 1.5 \times
10^{-4}$\,mag$^2$\,min).  This time sampling and photometric precision
is reachable with current planet transit surveys (e.g.,
Mall\'en-Ornelas et al. 2002). The transit shape must be well defined
by high photometric precision and high time sampling because most
star-planet parameters depend on transit shape. Specifically a
limiting factor is time sampling of ingress and egress and their start
and end times to determine transit shape.

\acknowledgements We thank Howard Yee, Tim Brown, Scott Gaudi, and
Gil Holder for useful discussions. S.S. is supported by the W.M.
Keck Foundation. S.S. thanks John Bahcall for valuable advice,
generous support, and encouragement. GMO thanks John
Bahcall and the IAS for generous hospitality during visits when
this work was carried out.

\end{document}